\documentclass[aps,prb,twocolumn,showpacs,amsmath,amssymb]{revtex4}
\usepackage{graphicx}
\usepackage{bm}

\newcommand{\eq}[1]{
\begin{equation}
{#1}
\end{equation}}

\begin{document}

\title{Direct access to quantum fluctuations through cross-correlation measurements}
\author{Iurii Chernii}
\author{Eugene V. Sukhorukov}
\affiliation{D\'epartement de Physique Th\'eorique, Universit\'e de Gen\`eve,
CH-1211 Gen\`eve 4, Switzerland}

\begin{abstract}

Detection of the quantum fluctuations by conventional methods meets certain obstacles, since it requires high frequency measurements.
Moreover, quantum fluctuations are normally dominated by classical noise, and are usually further obstructed by various accompanying effects such as a detector backaction. In present work, we demonstrate that these difficulties can be bypassed by performing the cross-correlation measurements. We propose to use a pair of two-level detectors, weakly coupled to a collective mode of an electric circuit.
Fluctuations of the current source accumulated in the collective mode induce stochastic transitions in the detectors. These transitions are then read off by quantum point contact (QPC) electrometers and translated into two telegraph processes in the QPC currents. Since both detectors interact with the same collective mode, this leads to a certain fraction of the correlated transitions. These correlated transitions are fingerprinted in the cross-correlations of the telegraph processes, which can be detected at zero frequency, i.e., with a long time measurements.
Concerning the dependance of the cross-correlator on the detectors' energy splittings $\varepsilon_1$ and $\varepsilon_2$, the most interesting region is at the degeneracy points $\varepsilon_1=\pm\varepsilon_2$, where it exhibits a sharp non-local resonance, that stems from higher order processes. We find that at certain conditions the main contribution to this resonance comes from the quantum noise. Namely, while the resonance line shape is weakly broadened by the classical noise, the height of the peak is directly proportional to the square of the quantum component of the noise spectral function.

\end{abstract}



\maketitle

\section{Introduction}

It has been recently understood that the noise phenomena are not necessarily something detrimental in physical experiments, but instead, they may carry a useful new information about the underlaying processes.\cite{beenakker schonenberger,blanter buttiker} This information may be difficult to extract from the measurements of average quantities, therefore properties of the fluctuations become themselves a valuable subject of research. Particularly, one may study higher irreducible moments (or cumulants) of physical quantities.\cite{levitov lesovik} It is especially interesting to detect essentially quantum properties of noise, such as non-symmetrized correlators, reflecting the quantum non-commutativity.\cite{RevQnoise}

To motivate this interest, it is useful to introduce an example of a quantum mechanical operator $j(t)$, representing a fluctuating electric current. Suppose this current is measured by a classical ammeter of certain bandwidth, and the raw measurement data is stored as a sequence of numbers $\langle j(t_k)\rangle$. The classical properties related to the symmetrized correlators, such as the noise power $S_{\rm sym}(\omega) = \int d\tau \exp(i\omega\tau) \langle \{\delta j(t),\delta j(t+\tau)\} \rangle$, or higher order symmetrized correlators, can be extracted from the raw data in postprocessing. By contrast, the quantum (anti-symmetrized) parts of the correlators can not be found in principle from postprocessing of such raw data, since this information is lost in the measurement of the average values $\langle j(t_k)\rangle$. Instead, these quantum properties may be inferred from the measurement of more complex quantities and systems, occurring naturally, or engineered on purpose. However, in this kind of measurements, it may not always be immediately clear which particular quantity is measured. Specifically, there is certain ambiguity concerning the questions of the operator ordering in the complex quantities, and of the process of their reduction to the classical values. Thus, a careful approach requires the knowledge of the detailed model of the detector and of the process of measurement.

\begin{figure}[b]
\centering
\includegraphics[width=0.4\textwidth]{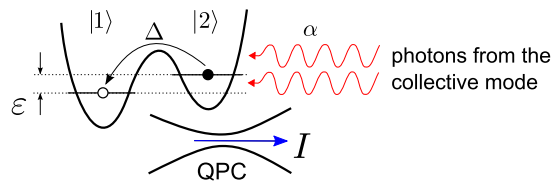}
\caption{A schematic representation of the two-level detector as a double well structure with the level splitting $\varepsilon$. Stochastic transitions in the detector are induced by the tunneling coupling $\Delta$ and assisted by photons coming from the collective mode.
The state of the detector is monitored by the QPC electrometer, located nearby one of the wells.}
\label{fig:detector}
\end{figure}

Hence one arrives at the notion of a mesoscopic on-chip detector.\cite{onchip1,onchip2,ESandAJ2,onchip3,ESandAJ1} It is a part of the measurement apparatus, which interacts directly with the system, and is responsible for transforming the quantum information into a classical signal. As an example, the two-level detectors have been studied theoretically and implemented in experiments,\cite{twolevel-aguado,twolevel-fujisawa,twolevel-khrapai,twolevel-gustavsson,twolevel-onac,twolevel-schoelkopf} and their operation is now quite well understood. Such a detector consists of a double quantum dot, or a similar structure, which has two energetically relevant quantum states (see Fig.\ \ref{fig:detector}), that have different spatial charge distributions. Then a noninvasive QPC electrometer \cite{qpc-field,q-measurements1,qpc-elzerman,qpc-fujisawa,qpc-schleser,qpc-vandersypen,charge-dicarlo,charge-fujisawa,sukhorukov1} may be used to read out the state of the two-level detector, and the resulting signal can then be amplified by conventional means.
In a properly adjusted operating regime, the two-level detector is weakly coupled to the mesoscopic system, and at short times they evolve together quantum mechanically. Due to the weakness of coupling, fluctuations in the system induce rare stochastic (nonadiabatic) transitions in the detector. Because of the noisy nature of these fluctuations, the state of the detector becomes decohered to a statistical mixture. \cite{q-measurements1,q-measurements5,q-measurements7,q-measurements8} In this case, the QPC electrometer effectively senses already classical state of the two-level detector, and thus, it actually satisfies our definition of an on-chip detector.

While the on-chip detector approach should at least clarify as what exactly is measured, the extraction of the information about the quantum fluctuations may still be challenging for some general reasons. The quantum effects often appear as small corrections to classical contributions, thus a high relative accuracy may be needed. Another source of complication is the fact that the system of interest is subjected to the perturbations induced by the measuring device itself,\cite{feedback-ensslin} and other extrinsic sources of noise. Therefore, some advanced techniques have to be employed to carefully extract the useful information.
One such technique, that demonstrated certain success at isolating the properties of the measured system is a so called cross-correlation technique. \cite{Reznikov2,Glattli} The main idea is that two (or probably more) detectors are used to measure certain fluctuating signal from the same system. Then, since the detectors are meant to be independent, any local processes at one of the detectors should not lead to cross-correlations in the fluctuations measured by different detectors. Therefore, the measurement of the cross-correlations of the detectors' outputs, gives certain level of protection from the local unwanted sources of noise, and from the detectors' backaction to the system, and thus may enhance the accuracy of the experiments.\cite{crosscorr1,crosscorr2,position1,position2,position3,jordan-buttiker}

\begin{figure}[b]
\centering
\includegraphics[width=0.33\textwidth]{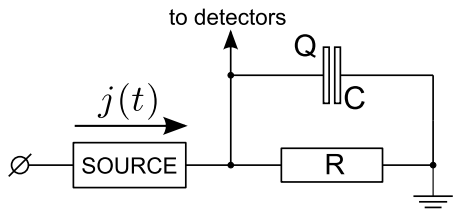}
\caption{The equivalent scheme of the measurement electric circuit coupling the noise source $j(t)$ to detectors via the collective mode $Q$.}
\label{fig:circuit}
\end{figure}

\begin{figure}[htb]
\centering
\includegraphics[width=0.42\textwidth]{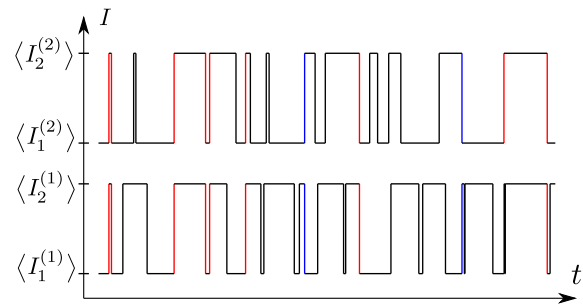}
\caption{ An example of the two telegraph processes generated by (partially) correlated switching of the two detectors. One of the mechanisms of cross-correlation is illustrated by correlated transitions $|11\rangle \rightleftarrows |22\rangle$ (red), and anti-correlated $|12\rangle \rightleftarrows |21\rangle$ (blue), both governed by the anti-diagonal elements of the transition matrix. The net cross-correlator may be estimated simply by counting the number of red lines, minus the number of blue lines.}
\label{fig:telegraph}
\end{figure}

In present work we consider using the cross-correlation technique to gain access to quantum fluctuations of current. We propose the following measurement setup: a mesoscopic system, source of current noise, is incorporated into an electric circuit, so the fluctuations of its current are accumulated on a capacitor (see Fig.\ \ref{fig:circuit}). An electric charge on the capacitor plays the role of a so-called collective mode, to which a pair of two-level detectors are coupled. We assume the weak coupling regime, when the evolution of the detectors' state can be described by the master equation for the average state occupations. Then fluctuations of the collective mode induce rare stochastic transitions in the detectors. These transitions generate two telegraph processes in the outputs of corresponding QPC electrometers, (see Fig.\ \ref{fig:telegraph}) and the cross-correlator of these outputs has to be measured in the markovian (long time) limit.

\begin{figure}[htb]
\centering
\includegraphics[width=0.45\textwidth]{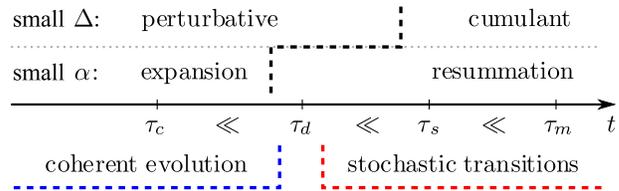}
\caption{A diagram of the hierarchy of the time scales in the measurement process: the noise correlation time $\tau_c$, the decoherence time $\tau_d$, switching time of the detectors $\tau_s$, and and the proposed measurement time $\tau_m$.}
\label{fig:timescales}
\end{figure}

Throughout the paper we rely on the concept of the time scale separation. In our model there are several independent small parameters (such as the coupling constant $\alpha$, and the tunneling amplitude $\Delta$) that produce the following grid of time scales (see Fig.\ \ref{fig:timescales}). The smallest scale is the noise correlation time $\tau_c$, at which the coherent quantum-mechanical evolution of the joint system of the noise source, the circuit, and the detector takes place. However, starting from the decoherence time $\tau_d \gg \tau_c$,  the noise source plays role of the heat bath for the detector. Then the evolution of the average occupations of the detector states is described by the master equation. These occupations vary on the characteristic time $\tau_s \gg \tau_d$ of the order of the inverse transition rates. Finally, the longest time scale $\tau_m \gg \tau_s$ is the markovian limit of the telegraph processes, where the cross-correlator should be measured.

To describe the quantum mechanical evolution of the joint system of the collective mode $\phi$ and the two-level detectors on short time scales $t < \tau_d$, it is rather more adequate to think of the two detectors in the space of four states $|11\rangle,|12\rangle,|21\rangle,|22\rangle$.
A calculation of the reduced density matrix for the detectors involves averaging over the fluctuations of the collective mode $\phi$. To the $n$th order of the perturbation expansion with respect to tunneling $\Delta$, this requires finding averages of the Keldysh ordered products of the corresponding number of exponential phase operators $e^{i\alpha\phi}$ (so called vertex operators). In the weak coupling regime $\alpha\ll1$, we use the cumulant expansion, since every next cumulant enters with one extra power of the small coupling constant, and we limit ourselves to the third cumulant. The main contribution comes from the times as long as the decoherence time $\tau_d$, where the cumulants can be expressed in terms of the zero-frequency expansion of the noise spectral functions. Quantum corrections to the classical long time asymptotic of the cumulants are small, and thus, they can be taken into account perturbatively in the coupling constant $\alpha$.

On the time scales $t>\tau_d$, we find the transition rates perturbatively in $\Delta$. We show that the most interesting effects do not appear on the level of the standard  $P(E)$ theory,\cite{ingold-nazarov,footnote-beyond-gaussian} which accounts tunneling to the lowest (second) order, only.
Notice, that some of the transitions between these states, such as $|11\rangle \rightleftarrows |22\rangle$ or $|12\rangle \rightleftarrows |21\rangle$, that arise in the 4th order in $\Delta$, directly correspond to simultaneous, correlated switching of the detectors and lead to cross-correlations at long times (see Fig.\ \ref{fig:telegraph}).
However, computation of the transition rates in the energy representation encounters well known divergences in higher orders of the perturbation theory. We avoid those divergences by considering directly the time evolution of the detector states. Namely, at times longer then the decoherence time $t>\tau_d$, but much shorter than the switching time $t\ll\tau_s$, the master equation description suggests linear in time drift of the occupations probabilities from the initial distribution. On the other hand, in the fourth order we find quadratic in time terms. We identify them with the reducible contribution generated in the perturbation expansion and, accordingly, find the transition rates by extracting the irreducible part from the long time asymptotic of the occupation probabilities.
As typical for the perturbation theory in higher orders, immense numbers of terms are generated. Nevertheless, we managed to find all our results analytically.\cite{footnote-mathematica}

Note, that the transition rates may be measured from the time-resolved observation of the telegraph processes.\cite{qpc-elzerman,qpc-fujisawa,qpc-schleser,qpc-vandersypen,charge-dicarlo,charge-fujisawa,sukhorukov1} Then, by fitting their dependence on the controllable parameters of the system, such as energy splittings and coupling constants, one may try to infer some of the noise properties. The drawback of this approach is that the large amounts of real-time data need to be recorded and analyzed. This also limits the possible measurement pace, since the real-time switching resolution is required to extract the transition rates. Instead, we expect that by measuring directly the cross-correlator of the two telegraph processes on the time scale $\tau_m\gg\tau_s$ longer than the switching time, one can considerably simplify the implementation of experiments.
To evaluate the resulting cross-correlator of the telegraph processes, we generalize for the case of two detectors the approach that has been proposed in Refs.\ [\onlinecite{FCS-ES Jordan}] and [\onlinecite{sukhorukov1}] to study the statistics of bistable systems. We present an exact general result, which is convenient to use if the transition matrix is symmetric, i.e., for the classical noise. For the case of quantum noise we also develop a perturbative in tunneling calculation which is better suitable for analytical computations.

Finally, we analyze the physically different contributions to the cross-correlator from classical and quantum noise.
Note, that although the two detectors are not directly interacting one with each other, an effective non-local interaction emerges between the detectors due to their coupling to the collective mode [see Eq.\ (\ref{H'})]. The effective interaction does not depend on the properties of the noise, but it leads to the trivial second order contribution to the cross-correlator. This contribution is not of much interest, since it can be activated by subjecting both, or even any one of the detectors to local classical noise. However, since it is proportional to the strength of the effective interaction, it can be minimized by tuning the parameters of the circuit.
By contrast, the more interesting kind of contributions arise as a consequence of the correlated response of the detectors to the fluctuations of the collective mode in the higher (fourth) order processes. The distinctive feature of these contributions is a sharp peak of the non-local resonance at the degeneracy point, where the detectors' level splittings satisfy $\varepsilon_1\pm \varepsilon_2 =0$.
We note, that the non-local resonance contains both classical and quantum parts of noise. Remarkably though, the classical part in the 4th order is also proportional to the effective interaction strength and thus may be rendered small, while the purely quantum contribution to the cross-correlator survives if the effective interaction is ``switched off''. The corresponding conditions (see Sec.\ \ref{sec:results analyzis}) can be achieved, basically, if the noise temperature is sufficiently high, and the coupling constants are made small enough.

The paper is organized as following: we present the model of two-level detectors in Sec.\ \ref{sec:detector model}.
Then, in Sec.\ \ref{sec: FCS} we give an overview of the counting statistics approach, generalize it for the cross-correlations, and present two methods of evaluating the cross-correlator, usable for classical and quantum noise respectively. In Sec.\ \ref{sec: cumulant expansion} and \ref{sec: correlation functions} we show how the cumulant expansion is applied to calculate the averages needed to find the transition rates.
Finally, in Sec.\ \ref{sec: results} we present the results for the classical and quantum noise cases, and analyze the conditions needed to access the required measurement regime.

\section{Two-level detectors \label{sec:detector model}}

An electric circuit may be modeled\cite{ingold-nazarov} by a set of bosonic fields, $\phi_k$, and their conjugated ``charges'' $q_k$.
We wish to single out one of these fields, a so called collective mode $Q$ and its phase $\phi$, which are linearly coupled to the rest of the fields and enter quadratically to the corresponding Hamiltonian of the circuit.
One can show that any such Hamiltonian may be transformed to the form, where all the couplings are carried by the phase $\phi$ only:
\eq{
\mathcal{H}_{c}= \frac{Q^2}{2C} + \mathcal{H}_{n}(\phi, q_k, \phi_k)
\label{circuit hamiltonian}
.
}
This model is sufficiently general to describe any circuit with the usual linear elements, such as resistors, capacitors and inductances, and can also include such mesoscopic elements as tunnel junctions by adding non-quadratic potentials to the Hamiltonian.

Let us assume that the collective mode $Q$ is linearly coupled, with a dimensionless coupling strength $\alpha$, to a two-level detector represented by the Hamiltonian
\eq{
\mathcal{H}_0 = \frac{\varepsilon}{2}\sigma_z + \Delta \sigma_x + \frac{\alpha}{2C}\sigma_z Q
,
}
where $\varepsilon$ is the energy level splitting, $\Delta \ll \varepsilon$ is a weak level mixing and we use the units $e=\hbar = 1$ throughout the paper.
The constant $\Delta$ can be also understood as the tunneling amplitude between the levels, or as the quantum level broadening.
This type of quantum detectors have been considered in a number of works and have been implemented experimentally.\cite{twolevel-aguado,twolevel-fujisawa,twolevel-khrapai,twolevel-gustavsson,twolevel-onac,twolevel-schoelkopf}

In what follows, we treat the term $\Delta\sigma_x$ as a smallest perturbation.
We, therefore, render the total Hamiltonian $\mathcal{H}=\mathcal{H}_0 + \mathcal{H}_c$ in a more convenient form by performing the following transformation:
\eq{
\mathcal{H}' = e^{ i  \frac{\alpha}{2}\sigma_z \phi} \mathcal{H} e^{- i  \frac{\alpha}{2} \sigma_z \phi}
.
}
This transformation affects only those terms that do not commute with $\sigma_z$ or $\phi$.
Since $[\phi,Q]=i$, it shifts the charge $Q$ by $ - \alpha\sigma_z /2$, canceling the linear coupling term  $\alpha \sigma_z Q / 2C$ and bringing interactions in the form of operators $e^{\pm i\alpha\phi}$,
\eq{
\mathcal{H}' = \frac{\varepsilon}{2}\sigma_z +
\Delta \! \left( \!
  \begin{array}{cc}
    0 & e^{i\alpha\phi} \\
    e^{-i\alpha\phi} & 0 \\
  \end{array} \!
\right)
- \frac{\alpha^2}{8C} + \mathcal{H}_c
,
}
where the energy is also shifted by a constant $- \alpha^2/8C$.
We switch to the interaction picture with the time dependent tunneling Hamiltonian
\eq{
\mathcal{H}_I(t) =
\Delta
\left( \begin{array}{cc}
 0 & e^{i[\phi(t)+\varepsilon t] } \\
 e^{-i[\phi(t)+\varepsilon t]} & 0 \\
  \end{array}
\right)
.
}
This suggests for a perturbative expansion in powers of $\Delta$. To justify this expansion, we assume that the $\Delta$ has to be the smallest energy scale in the system. Particularly, it must be smaller than the level broadening introduced by noise.

\subsection{Pair of two-level detectors}
Now, let us consider two such detectors, both coupled to the same circuit via the charge $Q$.
\eq{
\mathcal{H}_0= \!\! \sum_{j=1,2} \! \mathcal{H}_0^{(j)} \! = \!\! \sum _{j=1,2} \!\! \left\{\frac{\varepsilon_j}{2}\sigma^{(j)}_z \! + \Delta_j \sigma^{(j)}_x \! + \frac{\alpha_j}{2C}\sigma^{(j)}_z Q\right\}
,
}
where we denote quantities belonging to each one of the detectors with an additional index $j=1,2$.
In this case, the analogous transformation,
\eq{
\mathcal{H}' = e^{ \frac{i}{2} \sum_j \alpha_j\sigma^{(j)}_z \phi} \,\mathcal{H} e^{ - \frac{i}{2} \sum_j \alpha_j\sigma^{(j)}_z  \phi}
}
leads to
\begin{multline}
\mathcal{H}' =
\sum_{j=1,2} \Big\{
\frac{\varepsilon_j}{2}\sigma^{(j)}_z +
\Delta_j \!
\left( \!
  \begin{array}{cc}
    0 & e^{i\alpha_j\phi} \\
    e^{-i\alpha_j\phi} & 0 \\
  \end{array} \!
\right) - \frac{\alpha_j^2}{8C} \Big\}  \\
+ \mathcal{H}_c  + \  \frac{E_c}{2} \sigma^{(1)}_z \sigma^{(2)}_z
.
\label{H'}
\end{multline}
Thus, besides changing the energy by a constant, it also generates the cross-term with
\eq{
E_c = \alpha_1 \alpha_2 / 2C
\label{cross-term energy delta epsilon}
.
}
The cross-term is nothing but the Coulomb charging energy of the capacitor for the screened detector charges, and represents the effective non-local interaction between the detectors mediated by the circuit collective mode. At the degeneracy points $\varepsilon_1 =\pm \varepsilon_2$ this interaction leads to the quantum avoided-crossing level splitting of a value
\eq{
\Delta\varepsilon = 4\Delta_1\Delta_2 E_c / \varepsilon_1^2
\label{avoided-crossing level splitting energy}
.
}

Finally, after switching to the interaction picture, the tunneling Hamiltonian for two detectors takes the form
\begin{widetext}
\eq{
\mathcal{H}_I =
\left(
\begin{array}{cccc}
 0 & \Delta_2 e^{i \alpha_2 \phi + i (\varepsilon_2 - E_c)t }  & \Delta_1 e^{i \alpha_1 \phi + i (\varepsilon_1 - E_c)t  }  & 0 \\
 \Delta_2 e^{-i \alpha_2 \phi - i (\varepsilon_2 - E_c)t } & 0 & 0 & \Delta_1 e^{i \alpha_1 \phi + i (\varepsilon_1 + E_c)t } \\
 \Delta_1 e^{-i \alpha_1 \phi - i (\varepsilon_1 - E_c)t } & 0 & 0 & \Delta_2 e^{i \alpha_2 \phi + i (\varepsilon_2 + E_c)t } \\
 0 & \Delta_1 e^{-i \alpha_1 \phi - i (\varepsilon_1 + E_c)t } & \Delta_2 e^{-i \alpha_2 \phi - i (\varepsilon_2 + E_c)t } & 0
\end{array}
\right)
\label{tunneling Hamiltonian}
.
}
\end{widetext}
Note that if $E_c = 0$ the tunneling Hamiltonian (\ref{tunneling Hamiltonian}) reduces to a tensor sum $\mathcal{H}_I = \mathcal{H}^{(1)}_I \otimes \mathbb{E} + \mathbb{E} \otimes \mathcal{H}^{(2)}_I$. We will see that the presence of $E_c$ leads to a trivial mechanism of cross-correlations even in the presence of only local noise. However these cross-correlations vanish with small $E_c$, and thus may become dominated by some more interesting phenomena, such as effects of quantum fluctuations in higher order processes.

\subsection{Time evolution}
To study cross correlations in the detectors' output, we wish to consider the long-time limit, where evolution of the detectors' states can be described by the master equation
\eq{
\dot{\mathrm{P}} = \hat{M} \mathrm{P}
\label{master equation}
}
for the occupation probabilities $\mathrm{P}=(p_1,\ldots,p_4)$. Such description is valid if the decoherence time $\tau_d =(\alpha^2 R^2 S)^{-1}$ is much shorter, $\tau_d \ll \Delta^{-1}$, than the quantum time scale associated with the quantum level repulsion of a single detector, or $\tau_d \ll (\Delta\varepsilon)^{-1}$ in the degeneracy point. This conditions are equivalent to a requirement that the classical level broadening due to the noise is always stronger then the quantum avoided-crossing level repulsion.
Or, in other words, one may say that the stochastic switching time $\tau_s=1/(\Delta^2\tau_d)$ is longer than the dephasing time, $\tau_s\gg\tau_d$.

The time dependence of the reduced density matrix of the detectors is
\eq{
\rho(t)= \mathrm{Tr}_c \big[ U \tilde{\rho}(0) U^\dag \big]
\label{density matrix}
,
}
where
\eq{
U=\hat{\mathrm{T}}\exp{\left\{-i\int_0^t dt'\mathcal{H}_I(t')\right\}}
\label{evolution operator}
}
is the interaction picture evolution operator, the initial condition is represented by the full density matrix of the system of detectors together with the circuit $\tilde{\rho}(0) = \rho(0) \times \rho_c(0)$, and $\mathrm{Tr}_c $ means averaging over the circuit degrees of freedom.

As soon as $\rho(t)$ is found, one observes that the off-diagonal elements of $\rho$ decay exponentially over the time $\tau_d$. Therefore we can concentrate our attention on the probabilities
\eq{
p_k(t) = \rho_{kk}(t) = \mathrm{Tr}\Big[ \rho(0) \! \times \! \rho_c(0) \ U^\dag  |k\rangle \langle k| U \Big]
\label{p(t)}
,
}
where the trace is over all the degrees of freedom of the system.
The equation (\ref{p(t)}) may be recast in the form
\eq{
\mathrm{P}(t) = \big[\mathbb{E}+\hat{\mathcal{M}}(t)\big] \mathrm{P}(0)
\label{occupations at time t}
,
}
where the elements of the time dependent transition probability matrix $ \hat{\mathcal{M}}(t)$ read
\eq{
\Big\{\hat{\mathcal{M}}(t)\Big\}_{kl} = \mathrm{Tr}_c\Big[ \rho_c(0) \ \langle k| U^\dag  |l\rangle \langle l| U |k\rangle \Big] - \delta_{kl}
\label{M of t}
.
}
They may be found from (\ref{p(t)}) as the perturbative expansion $\hat{\mathcal{M}}(t) =~ \sum_n \hat{\mathcal{M}}_n(t)$ with respect to the tunneling Hamiltonian ~(\ref{tunneling Hamiltonian}).
Comparing the expression (\ref{occupations at time t}) with a solution of the equation (\ref{master equation}), $P(t)=\exp\{ \hat{M}t\} P(0)$, one finds that the master equation matrix can be obtained from the long time asymptotics of the irreducible part of $\hat{\mathcal{M}}(t)$
\eq{
\hat{M}=\lim_{t\rightarrow \infty} \frac{1}{t}\log\big[\mathbb{E}+\hat{\mathcal{M}}(t)\big]
\label{master equation matrix}
.
}

\begin{figure}[htb]
\centering
\includegraphics[width=0.37\textwidth]{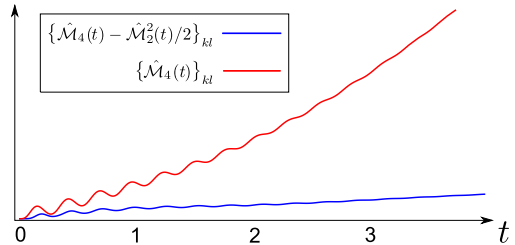}
\caption{Typical, quadratic in time dependence (red line) of the elements of bare $\hat{\mathcal{M}}_4(t)$, and linear in time dependence (blue line) of corrected $\hat{\mathcal{M}}_4(t)-\hat{\mathcal{M}}_2^2(t)/2$. Time is indicated in units of the decoherence time $\tau_d$.}
\label{reducible t2}
\end{figure}

Note, that due to the specific structure of (\ref{tunneling Hamiltonian}), only even powers of $\Delta_j$ are present in the diagonal elements of the perturbation series for the density matrix (\ref{density matrix}). Therefore, to the fourth order in $\Delta_j$, one can write $ \hat{\mathcal{M}}(t) = \hat{\mathcal{M}}_2(t) + \hat{\mathcal{M}}_4(t) $.
The time dependence of an arbitrary element of the $\hat{\mathcal{M}}_2(t)$ appears to be quite simple:
\eq{
\big\{ \hat{\mathcal{\mathcal{M}}}_2(t) \big\}_{kl} = C_0(t) + C_1 t
,
}
where $C_1$ is a constant, and the first term $C_0(t)$ is a decaying oscillating function of time. This decaying term only represents the artefact of the chosen decomposition of the initial condition $\tilde{\rho}=\rho \times \rho_c$, which, due to the detector-circuit interaction, does not describe a stationary state. However, the stationary, linear in time behavior is restored after the decoherence time $\tau_d$, as illustrated by the decaying oscillations in the Fig.\ \ref{reducible t2}.

The elements of the next term, $\hat{\mathcal{M}}_4(t)$, may have a more complex structure:
\eq{
\big\{ \hat{\mathcal{M}}_4(t) \big\}_{kl} = C'_0(t) + \big[C'_1+ D'_1(t)\big] t + C'_2 t^2
,
}
where the functions $C'_0(t)$ and $D'_1(t)$ are analogous to $C_0(t)$, the constant $C'_1$ is analogous to $C_1$, and there is a quadratic in time term $C'_2 t^2 $. This last term is a reducible part of the $\hat{\mathcal{M}}_4(t)$, which is exactly eliminated by the logarithm in (\ref{master equation matrix}).
Particularly, to the 4th order one has:
\eq{
\hat{M}=\lim_{t\rightarrow \infty} \frac{1}{t}\Bigl[\hat{\mathcal{M}}_2(t) + \hat{\mathcal{M}}_4(t)-\frac12 \hat{\mathcal{M}}_2^2(t)\Bigr]
\label{master equation matrix perturbative}
.
}
This expression provides a conclusive point in the calculation of the transition matrix $\hat{M}$ from the explicit time dependence of the occupations $P(t)$ .

\section{Full counting statistics of the QPC currents \label{sec: FCS}}

In this section we present the method for calculating the statistics of a QPC current using the generating functions approach.
We first recall the simpler case of a single detector \cite{FCS-ES Jordan,sukhorukov1} and later generalize it to the case of two detectors.
Consider a QPC as a charge detector tuned to detect the state of the two-level system (see Fig.\ \ref{fig:detector}). The statistics of the current passing through the corresponding conductance levels $k=1,2$ of the QPC is described at long times by the moment generating function
\eq{
g_k (\lambda,t)=\sum_n \exp(\lambda n) f_k(n) = e^{H_k(\lambda) t}
,
}
where $f_k(n)$ is the probability that $n$ electrons are transferred while the levels $k$ is occupied, $t$ is the total time of the measurement, and $H_k(\lambda)$ is the function generating current cumulants:
\eq{
\langle\!\langle I_k^m \rangle\!\rangle = \left. \frac{\partial^m H_k}{\partial \lambda^m}\right|_{\lambda \rightarrow 0}
.
}

One can see that $g_k (\lambda,t)$ satisfy the equations
\eq{
\dot{g}_k (\lambda,t)=H_k(\lambda) g_k (\lambda,t)
.
}
This equations may be modified in order to take into account mixing of the current channels induced by switching of the detector.
Introducing $\mathrm{G}(\lambda,t) \equiv (g_1, g_2)$, the extended master equation reads:
\eq{
\dot{\mathrm{G}}(\lambda,t) = \hat{W} \mathrm{G}(\lambda,t)
\label{extended ME}
,
}
where the matrix $\hat{W} = \hat{H} + \hat{M}$ is a sum of the cumulant generating functions matrix
\eq{
\hat{H} =
\left(
  \begin{array}{cc}
    H_1(\lambda) & 0 \\
    0 & H_2(\lambda) \\
  \end{array}
\right)
\label{sigle detector matrix H}
,
}
and the transition matrix
\eq{
\hat{M}  =
\left(
  \begin{array}{cc}
  - \Gamma_+  &  \Gamma_-  \\
   \Gamma_+  &  - \Gamma_- \\
  \end{array}
\right)
\label{sigle detector transition matrix}
,
}
where the transition rates are denoted as $\Gamma_\pm$. In the long time limit, the statistics of the QPC current is given by the generating function
\eq{
H(\lambda)= \lim_{t\rightarrow\infty}\frac{1}{t}\log \sum_k g_k (\lambda,t)
,
}
which is the largest eigenvalue of the matrix $\hat{W}$,
\eq{
H(\lambda) = H_s - \Gamma_s + \sqrt{(H_d + \Gamma_d)^2 + \Gamma_-\Gamma_+ }
\label{leading eigenvalue}
,
}
where $H_{s,d} = \frac{1}{2}(H_1 \pm H_2)$, and $\Gamma_{s,d} = \frac{1}{2}(\Gamma_+ \pm \Gamma_-)$.
Current cumulants are then found as derivatives of this generating function
\eq{
\langle\!\langle I^m \rangle\!\rangle = \frac{\partial^m H(\lambda)}{\partial \lambda^m}\Big|_{\lambda \rightarrow 0}
.
}

Taking into account that
\begin{multline}
\frac{\partial H(\lambda)}{\partial \lambda}\Big|_{\lambda \rightarrow 0} =
\sum_k
 \frac{\partial H}{\partial H_k}
\left. \frac{\partial H_k}{\partial \lambda} \right|_{\lambda \rightarrow 0} \\
 = \sum_k  \langle I_k\rangle  \left. \frac{\partial H}{\partial H_k} \right|_{H_k \rightarrow 0}
\label{composite derivative}
,
\end{multline}
we obtain the expected result for the average current:
\eq{
\langle I \rangle = \frac{\Gamma_- \langle I_1\rangle }{2\Gamma_s}  + \frac{\Gamma_+ \langle I_2\rangle }{2\Gamma_s}
= \bar{p}_1 \langle I_1\rangle  + \bar{p}_2 \langle I_2\rangle
,
}
and, similarly, for the noise
\eq{
\langle\!\langle I^2 \rangle\!\rangle =
\frac{\bar{p}_1 \bar{p}_2}{ \Gamma_s} \big(\langle I_2\rangle-\langle I_1\rangle\big)^2  + \bar{p}_1 \langle\!\langle I_1^2 \rangle\!\rangle + \bar{p}_2 \langle\!\langle I_2^2 \rangle\!\rangle
\label{auto-correlator}
.
}
Here $\langle I_k \rangle$ is the average current at the conduction level ~$k$ of the QPC, $\bar{p}_{2,1} =~\Gamma_\pm /(\Gamma_+ +\Gamma_-)$ are the stationary level occupations, and $\langle\!\langle I_k^2 \rangle\!\rangle$ is the corresponding zero frequency noise power of the QPC.

\subsection{Cross-correlator}

The above scheme can be easily generalized to a pair of two such detectors. Without level mixing, the joint probability distribution for the numbers of electrons passed through each QPC factorizes: $ f_{kl}(n_1,n_2) = f_{k}(n_1) f_{l}(n_2)$, therefore the corresponding generating functions acquire the form
\eq{
g_{kl}(\lambda,\eta,t) = g^{(1)}_k (\lambda,t)g^{(2)}_l (\eta,t)
,
}
with different counting variables $\lambda$ and $\eta$ for the two QPCs respectively, and satisfies the equation
\eq{
\dot{g}_{kl}(\lambda,\eta,t) = \Big[H^{(1)}_k(\lambda) + H^{(2)}_l(\eta) \Big] g_{kl}(\lambda,\eta,t)
,
}
where the upper index $(j)$ indicates the properties of the detector $j$.

Switching of the detectors is accounted for in the extended master equation, $\dot{\mathrm{G}} =~ \hat{W} \mathrm{G}$, with the matrix that has the same structure as before: $\hat{W}=\hat{M}+\hat{H}$, where $\hat{M}$ is the transition  matrix. In the basis of states $\bigl(|11\rangle,|12\rangle,|21\rangle,|22\rangle\bigr)$, the diagonal matrix $\hat{H}$ of the cumulant generating functions is:
\eq{
\hat{H} = \hat{H}_1\otimes \mathbb{E} + \mathbb{E} \otimes \hat{H}_2
,
}
where $\hat{H}_j$ are the single detector matrices (\ref{sigle detector matrix H})
\eq{
\hat{H}_j =
\left(
  \begin{array}{cc}
    H_1^{(j)} & 0 \\
    0 & H_2^{(j)} \\
  \end{array}
\right)
,
}
as functions of corresponding counting variables $\lambda$ or $\eta$.

As before, current cumulants may be found as derivatives of the largest eigenvalue $H(\lambda,\eta)$ of the matrix $\hat{W}$. Particularly, the current cross-correlator reads
\eq{
\langle\!\langle I_1 I_2 \rangle\!\rangle = \left. \partial_\lambda \partial_\eta H(\lambda,\eta) \right|_{\lambda,\eta=0}
\label{generating cross-correlator}
.
}
Since there is only one derivative with respect to each counting variable, the cumulant generators $H_{k}^{(j)}$ can be replaced by their corresponding average currents $H^{(1)}_k(\lambda) \approx  \lambda\langle I^{(1)}_k \rangle$ and $H^{(2)}_k(\eta) \approx \eta\langle I^{(2)}_k \rangle$.
Thus the problem of finding the cross-correlator of QPC currents can be reformulated as finding a second order correction to the largest eigenvalue of the matrix $\hat{M}$ from the perturbation $\hat{H}$.

Note, that due to the conservation of the total probability $\sum_j  \hat{M}_{jk}=0$, the transition matrix is degenerate, i.e., one of it's eigenvalues $m_0=0$, and all the others are negative, $m_{i\neq0} <0$. Therefore we can use the general result of the perturbation theory that $H(\lambda,\eta)$ can be found as a second order correction to the largest eigenvalue $m_0$:
\begin{multline}
H(\lambda,\eta) = \sum_{i\neq0} \frac{\langle m_0| \hat{H} |m_i \rangle  \langle m_i| \hat{H} |m_0 \rangle}{m_0 - m_i}  \\
 = -\langle m_0| \hat{H} \hat{M}^* \hat{H} |m_0 \rangle
 \label{pseudoinverse}
,
\end{multline}
where $ \hat{M}^{*}= \sum_{i\neq0} (1/ m_i) |m_i \rangle  \langle m_i|$ is also known as a Moore-Penrose pseudoinverse of the matrix $\hat{M}$; $|m_i \rangle$ are the eigenvectors\cite{footnote-left-right-eigenvectors} of $\hat{M}$.

To find the cross-correlator we need to calculate the derivative in Eq.\ (\ref{generating cross-correlator}). Since the counting variables only enter the matrix $\hat{H}$, it is convenient to introduce the following operator
\eq{
\hat{\mathcal{D}}=\frac{\partial^2}{\partial_\lambda \partial\eta} \left. \big(\hat{H} |m_0 \rangle \langle m_0| \hat{H} \big) \right|_{\lambda,\eta\rightarrow 0}
\label{proector matrix}
,
}
so that the cross-correlator may be expressed as following:
\eq{
\langle\!\langle I_1 I_2 \rangle\!\rangle = -\textrm{Tr}(\hat{\mathcal{D}} \cdot \hat{M}^{*})
.
}
It appears also, that the cross-correlator is always proportional to the differences of the current levels
$\Delta I_1 =~\!\!\langle I^{(1)}_2\rangle-~\!\!\langle I^{(1)}_1\rangle $  and
$\Delta I_2 = \langle I^{(2)}_2\rangle-\langle I^{(2)}_1\rangle $,
therefore it is natural to define a normalized cross-correlator, that has the dimensionality of time,
\eq{
X=\frac{\langle\!\langle I_1 I_2 \rangle\!\rangle}{\Delta I_1 \Delta I_2} = -\frac{\textrm{Tr}(\hat{\mathcal{D}} \cdot \hat{M}^{*})}{\Delta I_1 \Delta I_2}
\label{cross-correlator}
.
}

Finally, we note that in practice, for systems with sufficiently general transition matrix $\hat{M}$, the approach described above typically leads to cumbersome expressions, in particular, for the eigenvalues and the eigenvectors of $\hat{M}$. Therefore, we present a different way to evaluate the cross-correlator for a particular class of transition matrices.

\subsection{Perturbative approach \label{alternative approach}}

We wish to consider a particular case, when the effective interaction $E_c$ in the Hamiltonian (\ref{tunneling Hamiltonian}) can be neglected, $E_c \rightarrow 0$.
In this case one can easily find the cross-correlator perturbatively in tunneling amplitude $\Delta$.
The transition matrix can be represented as
\eq{
\hat{M} = \hat{M}_0 + \delta\!\hat{M}
\label{alternative transition matrix}
,
}
where $\hat{M}_0$ is evaluated to the 2nd order in tunneling and represents completely uncorrelated detectors, while $\delta\!\hat{M}$ is a small correction of the 4th order.
The matrix $\hat{M}_0 = \hat{M}_1\otimes \mathbb{E} + \mathbb{E} \otimes \hat{M}_2$ is a tensor sum of the single-detector transition matrices $\hat{M}_1$ and $\hat{M}_2$, defined according to Eq. (\ref{sigle detector transition matrix}) as
\eq{
\hat{M_j}  =
\left(
  \begin{array}{cc}
  - \Gamma^{(j)}_+  &  \Gamma^{(j)}_-  \\
   \Gamma^{(j)}_+  &  - \Gamma^{(j)}_- \\
  \end{array}
\right)
\label{sigle detector transition matriceS}
.
}
The extended master equation matrix then reads
\eq{
\hat{W} = \hat{M} + \hat{H} = \hat{W}_0 + \delta\!\hat{M}
,
}
where, the matrix $\hat{W}_0 = \hat{M}_0 + \hat{H}$ retains the same tensor structure:
$\hat{W}_0=  \hat{W}_1\otimes \mathbb{E} + \mathbb{E} \otimes \hat{W}_2$, and $\hat{W}_{1}$ and $\hat{W}_{2}$ are the single detector matrices as in (\ref{extended ME}).
The matrix $\hat{W}_0$ can be diagonalized explicitly, and it's largest eigenvalue is, obviously, the sum of the corresponding eigenvalues (\ref{leading eigenvalue}) for each detector separately, $H_0(\lambda,\eta) = H^{(1)}(\lambda) + H^{(2)}(\eta)$. Thus, as expected for the uncorrelated detectors,
\eq{
\langle\!\langle I_1 I_2 \rangle\!\rangle_0 = \partial_\lambda \partial_\eta H_0(\lambda, \eta) = 0
,
}
where the index $0$ stands for the lowest order contribution.
We see that any cross-correlations may only be generated by the correction matrix $\delta\!\hat{M}$, which brakes the tensor sum structure of $\hat{W}$.
\cite{footnote-transitions-interpretation}

As before, we need to find the largest eigenvalue $H(\lambda, \eta) = H_0(\lambda, \eta)+ \delta H(\lambda, \eta)$  of $\hat{W}$, where $\delta H(\lambda, \eta)$ is a correction due to the perturbation $\delta\!\hat{M}$.  Note, that in contrast to the general case, the counting variables are not contained in the perturbation $\delta\!\hat{M}$, but are absorbed into the matrix $\hat{W}_0$ (which makes $\hat{W}_0$ non-degenerate). Therefore, it is sufficient to use the first order perturbation theory:
\eq{
\delta H(\lambda, \eta) = \langle m_0 | \delta\!\hat{M} | m_0 \rangle
,
}
where $| m_0 \rangle$ is the eigenvector of $\hat{W}_0$ corresponding to the eigenvalue $H_0(\lambda, \eta)$.

Finally, the cross-correlator is given by
\eq{
\langle\!\langle I_1 I_2 \rangle\!\rangle = \left. \partial_\lambda \partial_\eta \delta H(\lambda,\eta) \right|_{\lambda,\eta=0}
,
}
and analogously to (\ref{composite derivative}), we find that
\eq{
\langle\!\langle I_1 I_2 \rangle\!\rangle =\sum_{k,l} \frac{\partial^2 \delta H}{\partial H^{(1)}_k\ \partial H^{(2)}_l}\Bigr|_{H^{(j)}_{k}=0}
 \langle I^{(1)}_k\rangle\langle I^{(2)}_l \rangle
\label{composite crossed derivative}
.
}
Although the full expression for $\delta H$ is somewhat cumbersome, it has an important property that the derivatives with respect to $H^{(1)}_k$ and $H^{(2)}_l$, all proportional to the same quantity $X_p$,
\eq{
\frac{\partial^2 \delta H}{\partial H^{(1)}_k \partial H^{(2)}_l}\Bigr|_{\lambda,\eta=0} = (-1)^{k+l} X_p
\label{normalization by Delta I}
.
}
Therefore, the cross-correlator (\ref{composite crossed derivative}) acquires the same form as in (\ref{cross-correlator}) and can be found using one of the relations (\ref{normalization by Delta I}):
\eq{
X_p =
\frac{2 \Gamma_{-}^{(1)} \Gamma_{+}^{(1)} \Gamma_{-}^{(2)} \Gamma_{+}^{(2)}}{(\Gamma_{-}^{(1)} + \Gamma_{+}^{(1)})^3 (\Gamma_{-}^{(2)} + \Gamma_{+}^{(2)})^3} \times \sum_{ij} \delta\!\hat{M}_{ij} \hat{\rm K}_{ij}
\label{alternative cross-correlator}
,
}
where to present a compact expression for $X_p$, we have introduced the matrix of the coefficients $\hat{\rm K}_{ij}$:
\begin{widetext}
\eq{
\hat{\rm K} =
\left(
\begin{array}{cccc}
 1 & \frac{\Gamma_{+}^{(2)}}{\Gamma_{-}^{(2)}} & \frac{\Gamma_{+}^{(1)}}{\Gamma_{-}^{(1)}} & \frac{\Delta\Gamma^{(1)} \Delta\Gamma^{(2)}}{2 \Gamma_{-}^{(1)} \Gamma_{-}^{(2)}}-1 \\
 -\frac{\Gamma_{-}^{(2)}}{\Gamma_{+}^{(2)}} & -1 & \frac{\Delta\Gamma^{(1)} \Delta\Gamma^{(2)} }{2 \Gamma_{-}^{(1)} \Gamma_{+}^{(2)}}+1 & -\frac{\Gamma_{+}^{(1)}}{\Gamma_{-}^{(1)}} \\
 -\frac{\Gamma_{-}^{(1)}}{\Gamma_{+}^{(1)}} & \!\! \frac{\Delta\Gamma^{(1)} \Delta\Gamma^{(2)} }{2 \Gamma_{+}^{(1)} \Gamma_{-}^{(2)}}+1 & -1 & -\frac{\Gamma_{+}^{(2)}}{\Gamma_{-}^{(2)}} \\
 \frac{\Delta\Gamma^{(1)} \Delta\Gamma^{(2)} }{2 \Gamma_{+}^{(1)} \Gamma_{+}^{(2)}}-1 & \frac{\Gamma_{-}^{(1)}}{\Gamma_{+}^{(1)}} & \frac{\Gamma_{-}^{(2)}}{\Gamma_{+}^{(2)}} & 1
\end{array}
\right)
,
}
\end{widetext}
and the notation $\Delta\Gamma^{(j)} = \Gamma_{+}^{(j)}-\Gamma_{-}^{(j)} $ is used, with $\Gamma_{\pm}^{(j)}$ being the transition rates to the second order of the perturbation theory.

\section{Cumulant expansion \label{sec: cumulant expansion}}
To find the transition rates (\ref{alternative transition matrix}), we expand the transition probabilities (\ref{M of t}) in the perturbation series with respect to the tunneling Hamiltonian (\ref{tunneling Hamiltonian}). We will henceforth use a simple angled brackets to denote the weighted average $\langle \ldots \rangle \equiv \mathrm{Tr}_c [\rho_c \ldots]$, which is used in (\ref{M of t}).
By substituting the evolution operator (\ref{evolution operator}), expanded perturbatively with respect to the tunneling Hamiltonian (\ref{tunneling Hamiltonian}), into the expression (\ref{M of t}) for the transition probability matrix, one can show that the elements of $\hat{\mathcal{M}}(t)$ may be expressed in terms of time integrals of the Keldysh ordered correlation functions of the following type:
\eq{
\Big\langle e^{i\gamma_1\phi_1} \ldots e^{i\gamma_n\phi_n} \Big\rangle =
\left. e^{F_n[\eta(t)]} \right|_{\eta(t)=\sum\limits_{j=1}^n i\gamma_j\delta(t-t_j)}
\label{cumulant expansion}
,
}
where $\phi_j \equiv \phi(t_j)$ for simplicity, the $\gamma_j$ take values of $\pm\alpha_1$ or $\pm\alpha_2$, and the generating function
\eq{
F_n[\eta(t)] = \sum_k  \frac{1}{k!} \! \int \! d^k t \prod_{j=1}^k \eta(t_j) \  \mathrm{T_K} \langle\!\langle \phi_1 \ldots \phi_k \rangle\!\rangle
}
is an expansions in terms of the Keldysh ordered cumulants $\mathrm{T_K} \langle\!\langle \phi_1 \ldots \phi_k \rangle\!\rangle$ of the order $k$.

We do not need to know all the correlators that appear in the expansion of the density matrix, but only those that contribute to the master equation through the matrices $\hat{\mathcal{M}}_2(t)$ and $\hat{\mathcal{M}}_4(t)$ from (\ref{master equation matrix perturbative}).
In the second order term $\hat{\mathcal{M}}_2(t)$, due to the structure of the tunneling Hamiltonian (\ref{tunneling Hamiltonian}), we find only the terms proportional to $\Delta_1^2$ or $\Delta_2^2$, thus ``originating'' separately from each detector.
In this case, we need the correlator of the form
\eq{
\langle e^{i\gamma\phi_1} e^{-i\gamma\phi_2} \rangle =
e^{-J^{(2)}_2 - i J^{(2)}_3}
\label{two phase cumulant expansion}
,
}
where the second and third cumulants, $J^{(2)}_2(t_1,t_2)$ and $ J^{(2)}_3(t_1,t_2)$, are found by expanding both sides of the equation (\ref{cumulant expansion}) in $\phi_j$:
\eq{
J^{(2)}_2(t_1,t_2) =
   \frac{\gamma^2}{2} \Big\langle \phi^2_1 - 2 \phi_1\phi_2 + \phi^2_2 \Big\rangle
\label{second order second cumulant}
,
}
\eq{
J^{(2)}_3(t_1,t_2) =
\frac{\gamma^3}{6} \Big\langle \phi^3_1 - 3 \phi^2_1 \phi_2 + 3 \phi_1 \phi^2_2 - \phi^3_2 \Big\rangle
\label{second order third cumulant}
.
}

By contrast, in the fourth order term $M_4(t)$, there is more diversity. First of all, there are terms proportional to $\Delta_1^2 \Delta_2^2$. Such terms are of our main interest, since they represent the non-local processes, involving both detectors.
Thus, we are interested in correlators
\eq{
\langle e^{i\gamma_1\phi_1} \ldots  e^{i\gamma_4\phi_4} \rangle =
e^{-J^{(4)}_2(t_1,\ldots,t_4) - i J^{(4)}_3(t_1,\ldots,t_4)}
\label{four vertex average}
,
}
where the four parameters $\gamma_j$ take four different values of $\pm\alpha_{1}$ and $\pm\alpha_{2}$.
Again, expanding both sides of the equation (\ref{four vertex average}), one finds
\eq{
J^{(4)}_2 = \frac12 \Bigl\langle \sum_k \gamma_k^2\phi_k^2 + 2 \sum_{k<l} \gamma_k \gamma_l \phi_k\phi_l \Bigr\rangle
\label{four point second cumulant}
,
}
and, in contrast with the result (\ref{second order third cumulant}), the third order cumulant
\begin{multline}
J^{(4)}_3 =
\frac{1}{6}\Big\langle  \sum_i \gamma_i^3 \phi_i^3 +3\sum_{i<j}\gamma_i \gamma_j \phi_i (\gamma_i\phi_i + \gamma_j\phi_j) \phi_j  \\
+ 6 \!\! \sum_{i<j<k} \gamma_i\gamma_j\gamma_k\phi_i \phi_j \phi_k \Big\rangle
\label{four point third cumulant}
\end{multline}
contains the three-point correlators $\langle \phi_i \phi_j \phi_k \rangle$ taken at three different times.

\section{Current and phase correlation functions \label{sec: correlation functions}}

From the equation of motion $\dot{\phi} = Q/C$ generated by the Hamiltonian (\ref{circuit hamiltonian}), it follows that the phase $\phi$ is an integral of the voltage on the capacitor. Since the constant component of voltage can be absorbed into the level splitting energy $\varepsilon$, we include only the fluctuations $\delta V$,
\eq{
\phi(t) = \int\limits_{-T}^{t} dt' \delta V(t')
,
}
where instead of integrating over time from minus infinity, we have introduced the regularization with some finite time $-T$, which is larger than any other timescale in the system.

The voltage fluctuations are related to the current fluctuations by the circuit impedance  $Z(\omega)$,
\eq{
\delta V(\omega) = Z(\omega) j(\omega)
.
}
Combining the last two equations together, one obtains
\eq{
\phi(t) = \int \frac{d\omega}{2\pi} Z(\omega) j(\omega) \frac{ e^{-i\omega t} - e^{i\omega T}}{(-i\omega)}
,
}
where, due to the regularization, a pole at $\omega =0$ is canceled by a zero in the numerator.
We further assume the limit of fast circuit, \cite{footnote-slow-circuit} i.e., the main contribution to the phase correlators comes from the times much longer than the circuit response time $\tau_{\rm RC} = RC$, and the impedance may be considered a constant $Z(\omega) = R$.

We define the current spectral functions $S_2(\omega)$ and $S_3(\omega_1,\omega_2)$ through a non-symmetrized two- and three-point current correlation functions
\eq{
\langle j(\omega_1)j(\omega_2) \rangle = 2\pi\delta(\omega_1 + \omega_2) S_2(\omega_1)
\label{spectral function s2}
,
}
\eq{
 \langle j(\omega_1)j(\omega_2)j(\omega_3) \rangle =
 2\pi\delta(\omega_1 + \omega_2 +\omega_3) S_3(\omega_1,\omega_2)
\label{spectral function s3}
,
}
where the delta-functions are the manifestation of the time translation invariance, since we consider a stationary process. Using these definitions, the phase correlators are expressed as following:
\begin{multline}
\langle \phi(t_1)\phi(t_2) \rangle = R^2 \!\! \int \!\! \frac{d^2\omega}{(2\pi)^2} \ \delta(\omega_1+\omega_2 ) \\
\times S_2 (\omega_1)  \prod_{j=1,2} \!\! \left (\frac{e^{-i\omega_j t_j}-e^{i\omega_j T}}{-i\omega_j} \right )
\label{two point phase correlation}
,
\end{multline}
and analogously,
\begin{multline}
\langle \phi(t_1)\phi(t_2)\phi(t_3) \rangle = R^3 \!\! \int \!\! \frac{d^3\omega}{(2\pi)^3} \ \delta (\omega_1+\omega_2 +\omega_3 ) \\
 \times S_3 (\omega_1,\omega_2) \prod_{j=1..3} \!\! \left (\frac{e^{-i\omega_j t_j}-e^{i\omega_j T}}{-i\omega_j} \right )
\label{three point phase correlation}
.
\end{multline}

Taking the integrals over $\omega$, we find the long time asymptotic behavior of the correlator (\ref{two point phase correlation}) as
\begin{multline}
\langle \phi(t_i)\phi(t_j) \rangle = R^2 \Bigl[ S\bigl(T + \min(t_i, t_j)\bigr)  \\ - i S' \ \textmd{sign}(t_i - t_j)\Bigr]
\label{two phase correlation with T}
.
\end{multline}
Here only the zero-frequency components of the spectral function contribute to the correlator:
\eq{
S_2(\omega) \simeq S + \omega S'
,
}
where $S\equiv S_2(0)$ is the classical noise power, and $S'\equiv ~dS_2(\omega)/d\omega|_{\omega=0}$ is the quantum part, representing non-commutativity of the current operator.

The correlator (\ref{two phase correlation with T}) has a divergent term proportional to~ $T$, which, however, does not contribute to any physically meaningful quantity. For instance, the cumulants (\ref{second order second cumulant}) and (\ref{four point second cumulant}) read:
\eq{
J_2(t_1,t_2) =
\frac{\alpha_j^2 R^2}{2} \Bigl(S|t_1 - t_2| - i S' \ \textmd{sign}(t_1 - t_2) \Bigr)
\label{second cumulant asymptotic}
,
}
while the cumulant from the fourth order can be rewritten in the similar form:
\begin{multline}
J^{(4)}_2 (t_1 \ldots t_4)
\\ = \frac{R^2}{2} \sum_{i<j} \alpha_i\alpha_j \Bigl(S|t_i - t_j| - i S' \ \textmd{sign}(t_i - t_j) \Bigr)
\label{second cumulant asymptotic 4th order}
.
\end{multline}
A weak coupling regime, considered throughout the paper, implies that either the detector to circuit coupling is weak (small $\alpha_j \ll 1$) or the impedance $R$ is small. Then the entire $J_2$ is small by the factor $\alpha_i\alpha_j R^2$, and the contribution to the time integrals of correlators (\ref{two phase cumulant expansion}) and (\ref{four vertex average}) that enter the perturbation expansion for $\hat{\mathcal{M}}(t)$, see Eq.\ (\ref{M of t}), comes from large differences of times of order $|t_i-t_j| \sim (\alpha_i\alpha_j R^2 S)^{-1}$. This justifies the use of asymptotic expressions of the type (\ref{second cumulant asymptotic}) and (\ref{second cumulant asymptotic 4th order}) and clarifies the  assumption of the fast circuit $ \tau_{\rm RC} \ll \tau_d=(\alpha^2 R^2 S)^{-1}$.

Analogously, for the three point correlator (\ref{three point phase correlation}), we expand the spectral function around zero frequencies,
\eq{
S_3(\omega_1,\omega_2) = S_3 + S_3^{(1)} \omega_1 + S_3^{(2)} \omega_2
,
}
and obtain
\begin{multline}
\langle \phi(t_1)\phi(t_2)\phi(t_3) \rangle = - R^3 S_3 \big[T+\min(t_j)\big]   \\
 - i R^3  \sum_{j=1,2} S_3^{(j)} \Bigl[ \theta\big(\min_{k\neq j}(t_k) - t_j\big) -\frac12 \Bigr]
.
\end{multline}
However, the parameters $S_3^{(1)}$ and $S_3^{(2)}$ are not independent. From the definition (\ref{spectral function s3}) one can find that $(S_3^{(1)})^* = S_3^{(1)} = 2 \textrm{Re} S_3^{(2)}$, therefore they are parameterized by two real numbers: the real and imaginary parts of $S_3^{(2)}$.
Again, although this correlators diverge with $T$, the third cumulant, such as (\ref{second order third cumulant}), is regular:
\begin{multline}
J^{(2)}_3 (t_1,t_2) =
  -\frac{\gamma^3 R^3}{6} \Bigl\{ S_3(t_1 - t_2) \\ + \frac{3i}{2} \big[ \mathrm{Re} S_3^{(2)} + i \ \mathrm{Im} S_3^{(2)}\mathrm{sign}(t_1-t_2) \big] \Bigr\}
\label{third cumulant asymptotic 2nd order}
.
\end{multline}
Note, that in contrast to the second cumulant (\ref{second cumulant asymptotic}), which depends on the absolute value $|t_1 - t_2|$, the classical part of the third cumulant depends on the difference $(t_1 - t_2)$, and because of this, it enters the transition rates the same way as the energies $\varepsilon_j$. This leads to a specific shift of the energies, discussed later. A similar expression is obtained for the third cumulant (\ref{four point third cumulant}) to the fourth order, but it is not displayed here, because it is somewhat cumbersome.

\section{The cross-correlator: Results \label{sec: results}}

Using the cumulant expansions such as (\ref{second cumulant asymptotic}), (\ref{second cumulant asymptotic 4th order}), and (\ref{third cumulant asymptotic 2nd order}), we can find the correlators (\ref{two phase cumulant expansion}) and (\ref{four vertex average}), which we need to calculate the transition probability matrix (\ref{M of t}). From the latter, we extract the master equation matrix (\ref{master equation matrix perturbative}) and, finally, find the cross-correlator using one of the results (\ref{cross-correlator}) or (\ref{alternative cross-correlator}).
We assume that the cross-correlator is a sufficiently smooth function of the parameters of the system, and thus, consider separately different contributions, each being suppressed by some small parameter. This allows us to estimate ratios of these contributions and determine regions of the parameters where one or another contribution is dominating.
We start with the case when the effective interaction $E_c$ in the Hamiltonian is neglected, and the noise is classical. Then the detector outputs become completely uncorrelated. We consecutively ``turn ~on'' the effective interaction and the quantum parts of the noise cumulants to investigate separately each contribution.

\subsection{Classical noise}

For the classical noise, the current operator $j(t)$ may be replaced by a classical variable, which commutes with itself in different moments of time. Therefore, the spectral function $S_2(\omega)$ becomes a symmetric function, and it's derivative at zero frequency is $S'=0$. In the same way, the three-point spectral function (\ref{spectral function s3}) has zero slope $S_3^{(1)}=S_3^{(1)}=0$, and we set $S_3(\omega_1,\omega_2)=S_3$.

The transition matrix $\hat{M}$ for the classical noise is symmetric. Taking into account that the sum over columns of the elements of $\hat{M}$ is zero, one can show that the stationary solution of the master equation, corresponding to the largest eigenvalue, is $|m_0\rangle = \frac14(1,1,1,1)$ with equal occupations of $p_j=1/4$ for all the four states, which can be regarded as a limit of high temperature. The matrix (\ref{proector matrix}) in this case is a constant independent on the parameters of the system, and the pseudoinverse $\hat{M}^{*}$ can be found explicitly. Note, that since the matrix $\hat{M}$ is symmetric, it is parametrized by 6 elements of its upper-right triangle. Then, finally, the classical contribution to the cross-correlator is expressed in terms of those elements:
\begin{multline}
X_{\rm cl} = \frac{1}{ 32{\textstyle \sum_{2} } }  \\
\times \Big[(m_{12}\! - m_{34}) (m_{13}\! - m_{24}) - (m_{14}\! - m_{23}){\textstyle \sum_{1}\nolimits }\Big]
\label{classical XC}
\end{multline}
where ${\textstyle \sum_{1} } = m_{12} + m_{13} + m_{24} + m_{34}$ , and the sum in the denominator  $\sum_{2} = \sum m_{ij} m_{kl} m_{nm}$ is over products of all combinations of three different elements from the 6 independent elements of $\hat{M}$.

\begin{figure}[htb]
\centering
\includegraphics[width=0.45\textwidth]{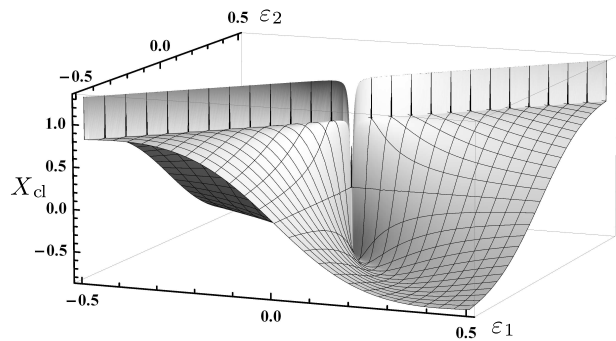}
\caption{The classical contribution $X_{\rm cl} = X^{(2)}_{\rm cl}+X^{(4)}_{\rm cl}$ to the cross-correlator in units of the dephasing time $1/\Omega = (\alpha_1 \alpha_2 R T_{\rm eff})^{-1}$ as a function of the detector level splittings $\varepsilon_1$ and $\varepsilon_2$ in units of the effective temperature $T_{\rm eff}$, at $\Delta_j=0.07\Omega, \alpha_j=0.55\pm0.05, R=0.12$, and $C=0.8/\Omega$. Note that the second order part $X^{(2)}_{\rm cl}$ does not exhibit any singularities at the non-local resonances $\varepsilon_1 \pm \varepsilon_2 \rightarrow 0$, and the sharp peaks come from the fourth order part $X^{(4)}_{\rm cl}$. }
\label{fig:classical}
\end{figure}

The expression (\ref{classical XC}) gives certain interpretation of how different combinations of the transition rates contribute to the cross-correlator, particularly, if the transition rates in one detector depend on the current state of the other detector. For example, if $m_{12} >  m_{34}$, $m_{13} > m_{24}$, then the positive cross-correlations are induced.
One important observation is that if the effective interaction term in the Hamiltonian is neglected, then the transition matrix acquires additional symmetry $m_{ij}=m_{kl}$ for $i+l=j+k=5$ (symmetry relative to the anti-diagonal), and $m_{14}=m_{23}$. Then the cross-correlator is identically equal to zero. In other words, the cross-correlator for classical noise vanishes together with the effective interaction strength $E_c$.

If the transition rates are calculated only to the second order in $\Delta_j$, the anti-diagonal elements of $\hat{M}$ are zero, and the result (\ref{classical XC}) simplifies further. Substituting corresponding transition rates, we obtain the classical contribution, shown in the Fig.\ \ref{fig:classical}:
\begin{multline}
X^{(2)}_{\rm cl} =   \frac{1}{4R^2 S} \\
\times \!
\frac{\varepsilon_1 \varepsilon_2 E_c^2}{\Delta_2^2 \alpha_2^2 \big(\varepsilon_1^2+E_c^2+\Omega_1^2 \big) \ + \ \Delta_1^2 \alpha_1^2 \big(\varepsilon_2^2+ E_c^2+ \Omega_2^2\big) }
\label{X cl exact}
,
\end{multline}
where we introduce the classical level broadening,
\eq{
\Omega_j = \alpha_j^2 R^2 S / 2
,
}
and recall that $E_c$ is the effective non-local interaction energy (\ref{cross-term energy delta epsilon}).
The typical maximum magnitude of this quantity, that is acquired at $\varepsilon_1/\varepsilon_2 \sim \pm \alpha_1\Delta_1 \big/ \alpha_2\Delta_2$, can be estimated as
\eq{
X^{(2)}_{\rm cl} \approx \pm \frac{E_c^2}{8 \alpha_1\alpha_2\Delta_1\Delta_2 R^2 S} = \pm \frac{E_c^2}{16 \Delta_1\Delta_2 \sqrt{\Omega_1 \Omega_2}}
\label{X bg}
.
}
It contains competing small parameters: $\Delta_j$ that are considered the smallest energy scales in the system, and effective interaction strength $E_c$.
It is especially remarkable that $E_c$ is a property of the circuit itself and does not depend on the source of noise. In fact, the quantity $X^{(2)}_{\rm cl}$ reflects rather the effect of the effective non-local coupling, than of the common noise source. If we replace the common noise source with two independent local sources, or even only one local source of noise, the cross correlations will persist.
For the same reason, this classical contribution (\ref{X cl exact}) is a well bounded function of the energies $\varepsilon_{1}$ and $\varepsilon_{2}$, and it does not exhibit singularities at the degeneracy points $\varepsilon_1 = \pm \varepsilon_2$.

By contrast, if the transition rates are calculated up to the fourth order in $\Delta_j$, then we obtain a correction $X^{(4)}_{\rm cl}$, which manifests resonance peaks at $\varepsilon_2= \pm \varepsilon_1$, that become more pronounced as the coupling constants approach each other. Naturally, this is the most interesting region of parameters. Using notations $\alpha_{1,2}=~\alpha \pm \delta\alpha/2$, $\varepsilon=~(\varepsilon_1 + \varepsilon_2)/2$, and $\Omega_{1,2}\sim\Omega=\alpha^2 R^2 S/2$, we present the asymptotic form of the fourth order classical contribution to the cross-correlator
\eq{
X^{(4)}_{\rm cl} \approx  \frac{ \Omega_{12} E_c^2}{16 \Omega^2 \big[ (\varepsilon_2 - \varepsilon_1)^2 + \Omega_{12}^2 \big]}
,
}
where
\eq{
\Omega_{12} = \frac12 (\alpha_1-\alpha_2)^2 R^2 S
}
is the classical broadening of the non-local resonance. Here we assumed that $\Omega_{12}$ is larger than the avoided-crossing energy splitting $\Delta\varepsilon$ and the classical level broadenings due to the local sources of noise.
The height of the peak is equal to
\eq{
\max X^{(4)}_{\rm cl} \approx \frac{E_c^2}{16 \Omega_{12} \Omega^2}, \ \ \textrm{at} \ \varepsilon_1 = \varepsilon_2
\label{X cl}
.
}

\subsection{Quantum noise}

In this section we concentrate on the effects of quantum noise, i.e., we now take into account the antisymmetric parts of the spectral functions $S_2(\omega)$ and $S_3(\omega_1,\omega_2)$. At the same time, we set $E_c=0$, so the tunneling Hamiltonian (\ref{tunneling Hamiltonian}) becomes a tensor sum, as well does the second order part of the transition matrix, $\hat{M}_2$, but not the fourth order correction $\hat{M}_4$. This allows us to obtain analytic results using the approach described in the section (\ref{alternative approach}), where we associate $\hat{M}_2$ with $\hat{M}_0$ in (\ref{alternative transition matrix}), and the fourth order part $\hat{M}_4$ with the perturbation $\delta\!\hat{M}$.

\begin{figure}[htb]
\centering
\includegraphics[width=0.45\textwidth]{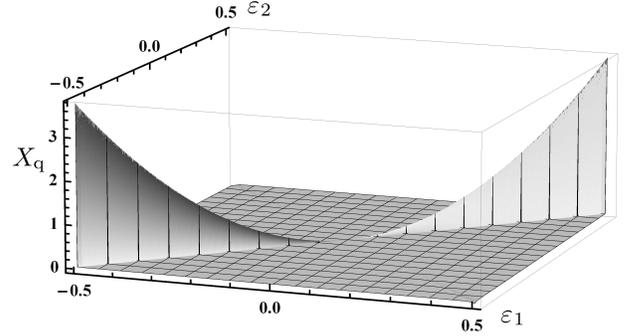}
\caption{Cross-correlations due to the quantum Gaussian component of noise over the classical background displayed in Fig.\ \ref{fig:classical}, in the same units, as a function of the energies $\varepsilon_{1}$ and $\varepsilon_{1}$ at $\alpha_1, \alpha_2 >0$, develop a sharp resonance peak on the diagonal, the height of which is proportional to the energy  squared.}
\label{fig:gaussian}
\end{figure}

In the limit of weak coupling, the main contribution from the correlator (\ref{two point phase correlation}) to the transition rates comes from long times. This means that the first term in the cumulant, $S |t_1 - t_2|$, is much larger then the second term $i S' \textrm{sign}(t_1 - t_2) $. This allows to include the quantum corrections perturbatively. Since the third cumulant contains one more coupling constant then the second cumulant, we expect the main contribution to come from Gaussian fluctuations.
Surprisingly, however, the first order contribution of $S'$ vanishes, and we have to keep the expansion at least up to $S'^2$. The fact that the Gaussian fluctuations do not contribute to the cross-correlator in the lowest possible order with respect to coupling constant means that we have to check if the contribution of third order correlations may enter with a lower power of the coupling constant. Unfortunately, this is not the case, and non-Gaussian correlations only make a small correction to the Gaussian contribution, as both of them enter quadratically into the cross-correlator (\ref{quantum cross-corr}).
As a function of energy splittings, $X_{\rm q}$ exhibits the non-local resonances at $\varepsilon_1 = \pm \varepsilon_2$. The exact expression is somewhat cumbersome, thus we do not display it, but concentrate on one of the resonances and find the asymptotic form:
\begin{multline}
X_{\rm q} \approx \frac{\varepsilon^2 \Omega_{12}}{64 S^2 }  \\ \times
\frac{ 16(S')^2 +  \big[4 \alpha^2 + 15  (\delta\alpha)^2 \big] R^2 \big(\mathrm{Re} S_3^{(2)}\big)^2}{(\varepsilon_1-\varepsilon_2)^2 + \Omega_{12}^2 }
\label{quantum cross-corr}
.
\end{multline}
Note, that in contrast to Eq.(\ref{X cl}) the hight of the resonance in the present case
\eq{
\max X_{\rm q} \approx  \frac{\varepsilon^2 (S')^2}{4 \Omega_{12} S^2}
\label{X q}
}
is proportional to the square of the energy $\varepsilon$, as can be seen in Fig.\ \ref{fig:gaussian}.

To complete this subsection, let us emphasize, that the resonant quantum contribution to the cross-correlator is an essentially non-local effect. In the limit when the effective interaction $E_c$ is rendered vanishingly small, the quantum contribution becomes the only one present, that carries the manifestation of the non-local quantum correlation.

\subsection{Analysis of physical conditions for the quantum noise detection. \label{sec:results analyzis}}
One may notice that in the results presented above in this section we have not included the classical non-Gaussian noise $S_3$. This is because the effect of this component of noise has been discussed earlier\cite{Sukhorukov2} and this effect is easy to describe: it consists solely in a uniform shift of the energies $\varepsilon_j \rightarrow \tilde{\varepsilon}_{j}  = \varepsilon_{j} + \alpha^3_{j}R^3 S_3/6$, and, plus to that, an additional small shift of $3\alpha_1\alpha_2 \delta\alpha R^3 S_3$ applies to the energy position of the non-local resonance  $\tilde{\varepsilon}_2 - \tilde{\varepsilon}_1$. Obviously, this shifts become irrelevant if we are interested in the regions of larger energies $\varepsilon_j \gg \Omega_j$.

\begin{figure}[htb]
\centering
\includegraphics[width=0.4\textwidth]{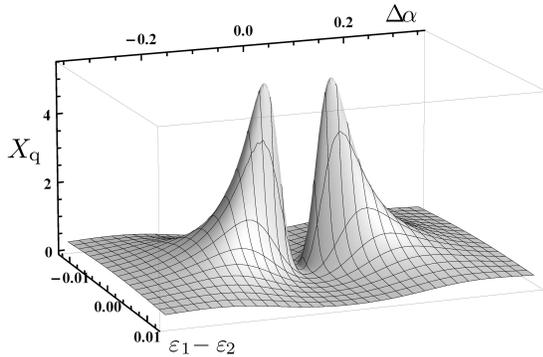}
\caption{The cross-section of the resonant peak of the quantum contribution to the cross-correlator as a function of $\delta\alpha=\alpha_2 - \alpha_1$  and the energy difference $\varepsilon_2-\varepsilon_1$ illustrates how the peak becomes thinner and, finally, disappears as $\delta\alpha \rightarrow 0$. The gap width is determined by the local source noise, that also limits the height of the resonance. Same units as in previous figures are used.}
\label{fig:resonance}
\end{figure}

Another effect that we have studied, but that was, for the sake of simplicity, omitted until now, is the influence of the local noise sources. These local sources may be modeled by an additional fluctuating phases $\beta_j \phi_j$ added to the collective mode phase $\alpha_j \phi$, including respective coupling constants $\beta_j$. Introducing corresponding noise powers $\beta_j^2  S^{(j)}_{\rm loc}$, one can find, that at the local resonances, these will simply add up to the classical level broadening $\Omega_j \rightarrow \Omega_j + \beta_j^2  S^{(j)}_{\rm loc}$.
More important, however, is their contribution to the classical level broadening of the non-local resonance: $\Omega_{12} \rightarrow \Omega_{12} + \sum_j \beta_j^2 S^{(j)}_{\rm loc}$. Thus, if we wish that this does not significantly wash out the resonance, then, besides of requirement of weakness of the local noise $\beta_j^2  S^{(j)}_{\rm loc} \ll \Omega_j$ , also the difference $\delta\alpha=\alpha_2 - \alpha_1$ must be kept large enough.  The effect of local noise is best seen in the Fig.\  \ref{fig:resonance} as a finite width gap between the two peaks at $\delta\alpha >0$ and at $\delta\alpha < 0$. Certain optimal value of the coupling constants may be chosen, that maximizes the magnitude of the non-local resonance.

We have to determine if there is a range of physical parameters, where the quantum contribution (\ref{X q}) to the cross-correlator is greater then the classical resonance contribution (\ref{X cl}) and the classical background (\ref{X bg}). In order to do this, we concentrate again on the non-local resonance $\varepsilon_j=\varepsilon$ where the quantum effects are maximal. We recall the expression for the effective non-local interaction strength (\ref{cross-term energy delta epsilon}), and our assumptions, made throughout the paper: (i) the level mixing constant is smaller than the the classical level broadening $\Delta \ll \Omega_j$; (ii) the circuit is fast $\tau_{\rm RC} \ll (\Omega_j)^{-1}$; and (iii) the zero-frequency expansion for the spectral functions is valid for energies smaller than the noise temperature $\varepsilon < T_{\rm eff}$.
Demanding that the quantum contribution is larger then the classical background, $X_q>X^{(2)}_{\rm cl}$, and using (iii), we obtain the inequalities
\eq{
T_{\rm eff} > \varepsilon > \frac{\delta\alpha}{\alpha} \frac{E_c}{\Delta}  T_{\rm eff}
,
}
where we have substituted $S/S' = 2 T_{\rm eff}$.
These inequalities are consistent when $E_c/\Delta < 1$ or
\eq{
\alpha^2 R  < \Delta \tau_{\rm RC} \ll 1
\label{consistent condition}
,
}
i.e., if the coupling is made sufficiently weak. However, since the background is smooth, one can demand that the quantum contribution is larger than only the resonant part of the classical contribution, $X_q>X^{(2)}_{\rm cl}$. This requirement leads to the condition
\eq{
T_{\rm eff} > \varepsilon > \frac{E_c}{\Omega}  T_{\rm eff}
,
}
which is weaker then (\ref{consistent condition}), since $\Delta \ll \Omega$. This condition may also be reformulated as
\eq{
T_{\rm eff} \tau_{\rm RC} >1
.
}
Hence one can see, that although $\tau_{\rm RC} \ll \Omega_j^{-1} = 1/(\alpha^2 R T_{\rm eff})$, the above inequality can be easily satisfied, in the weak coupling limit $1/\alpha^2 R \gg 1$.

Finally, we can also compare the cross-correlator with one of the auto-correlators $A_j=~\langle\!\langle I^2_j \rangle\!\rangle / (\Delta I_j)^2 $, which can be found from (\ref{auto-correlator}) and estimated as inverse of the typical transition rate, far from the local resonance $ A_j \approx \varepsilon_j^2/(\Delta_j^2 \Omega_j) $. Thus, even in the resonance, the cross-correlator to auto-correlator ratio, $X_q/A_j \sim \alpha^6 R^2/(\delta\alpha)^2 \ll 1$, is still much smaller then one. In other words,  correlated transitions make only a small fraction of all transitions.

\section{Summary and conclusions}

Motivated by the challenges on the way of measuring the properties of quantum fluctuations, we have presented in detail one possible approach to the problem, based on the cross-correlation technique. We use a concept of the collective mode, through which fluctuations of the current of the noise source are transferred to the on-chip detector. The latter consists of a pair of the two-level systems, monitored by QPC charge detectors. Thus, noise induced transitions generate two telegraph processes in the QPC currents, cross-correlator of which is supposed to be measured.
The cross-correlation technique allows one to avoid the effects of local noise sources and can provide direct access to the quantum component of the collective mode noise.

In the weak tunneling regime, the switching of the detectors can be described in terms of the master equation. The cross-correlator, on the other hand, is found using the extended master equation formalism, which describes the evolution of the probability distribution in the mixed space: for the charge, transmitted through the QPCs, and the occupations of the two detectors' states. Corresponding generalized transition matrix includes both the rates of the transitions of the detectors and the cumulant generators of the currents in the QPCs. The cross-correlator is then obtained from the largest eigenvalue of the generalized transition matrix. We propose two methods of the evaluation of this eigenvalue, each method having its own advantages for different symmetries of the transition matrix in the cases of the classical and quantum noise.

To find the transition rates we calculate directly the time evolution of the detectors' state to the fourth order of the perturbation theory. While, typically, this leads to well known divergences originating from the continuum of the environmental states, we avoid those divergences in a well controlled way by using the idea of the separation of time scales.
We consider a time interval much shorter than a characteristic time of the evolution of the detectors, and much longer than the correlation time of the noise, that induces transitions. On this time interval, the solution of the master equation suggests a linear in time variation of the occupation probabilities, while the formal perturbation expansion of the transition probabilities, with respect to the tunneling Hamiltonian, gives rise to the quadratic in time terms. We observe that the quadratic in time terms are reducible, and thus cancel by taking the logarithm of the time dependent transition matrix. This leads to the final results for the elements of the transition matrix.

By turn, perturbative calculation of the time evolution of the transition probabilities requires averaging the products of two and four vertex operators over the fluctuations of the collective mode.
In the limit of weak coupling of the detectors to the collective mode, the cumulant expansion has been used, and the cumulants are expressed through the components of the zero-frequency expansion of the current noise spectral functions. Such an expansion is justified in the limit of weak coupling, because the main contribution to time integrals for the elements of the transition matrix comes from long times.
Note that the weak coupling resummed in this way, leads to the classical broadening of the quantum resonances at $\varepsilon_j=0$ in the transition rates of each detector separately. But due to the local nature of these resonances, they do not contribute to the cross-correlator.

However, interaction of the detectors with the collective mode leads to the effective non-local coupling between the detectors. Already to the 2nd order in tunneling, this effective coupling contributes to the cross-correlator a smooth background, which is induced by the classical noise, and thus does not represent interest.

Finally, a more interesting effect appears in the 4th order of the perturbation theory with respect to tunneling Hamiltonian. Namely, at the degeneracy points $\varepsilon_1\!=\pm \varepsilon_2$ of the two detectors, a narrow non-local resonance arises. We show that when the effective coupling between two detectors is reduced, the main contribution to this non-local resonance arises from the quantum component of the noise source current.
We estimate this effect, compare it to classical contribution to the cross-correlator and propose the range of parameters where the quantum contribution dominates. We show that the observation of quantum noise is feasible in the regime of weak coupling. Moreover, we argue that a sharp non-local resonance from the quantum noise may be well visible over the smooth background of the classical contribution, and has a characteristic energy dependence that can be used to distinguish it from the classical resonance.

\begin{acknowledgments}
We thank Ivan P. Levkivskyi for fruitful discussions.
This work was supported by the Swiss National Science Foundation.
\end{acknowledgments}


\begin{thebibliography}{9}


\bibitem{beenakker schonenberger}
C.W.J. Beenakker, C. Schonenberger,
Physics Today {\bf 56} (5), 37-42 (2003)

\bibitem{blanter buttiker}
Ya. M. Blanter, M. B\"uttiker ,
Phys. Rep. {\bf 336}, 1 (2000).


\bibitem{levitov lesovik}
L.S. Levitov, H. Lee, and G.B. Lesovik,
J. Math. Phys. {\bf 37}, 4845 (1996).


\bibitem{RevQnoise}
A. A. Clerk, M. H. Devoret, S. M. Girvin, Florian Marquardt, R. J. Schoelkopf
Rev. Mod. Phys. {\bf 82}, 1155–1208 (2010).

\bibitem{onchip2}
T.T. Heikkil\"a, P. Virtanen, G. Johansson, F.K. Wilhelm,
Phys. Rev. Lett. {\bf 93}, 247005 (2004).

\bibitem{onchip3}
J. Ankerhold, and H. Grabert,
Phys. Rev. Lett. {\bf 95}, 186601 (2005).


\bibitem{ESandAJ1}
E.V. Sukhorukov and A.N. Jordan, Phys. Rev. Lett. {\bf 98},
136803 (2007).


\bibitem{ESandAJ2}
A. N. Jordan, and E. V. Sukhorukov,
Phys. Rev. B {\bf 72}, 035335 (2005)


\bibitem{onchip1}
E. B. Sonin, Phys. Rev. B {\bf 70}, 140506 (2004).



\bibitem{twolevel-fujisawa}
T. Fujisawa, T.H. Oosterkamp, W.G. van der Wiel,
B.W. Broer, R. Aguado, S. Tarucha, and L.P. Kouwenhoven,
Science {\bf 282}, 932 (1998).

\bibitem{twolevel-aguado}
R. Aguado and L.P. Kouwenhoven,
Phys. Rev. Lett. {\bf 84}, 1986 (2000).

\bibitem{twolevel-schoelkopf}
R.J. Schoelkopf, A.A. Clerk, S.M. Girvin, K.W. Lehnert, and M.H. Devoret,
in {\em Quantum Noise in Mesoscopic Physics}, edited by Y.V. Nazarov
(Kluwer, Dordrecht, 2003).

\bibitem{twolevel-onac}
E. Onac, F. Balestro, L.H. Willems van Beveren, U. Hartmann, Y.V. Nazarov, and L.P. Kouwenhoven,
Phys. Rev. Lett. {\bf 96}, 176601 (2006)

\bibitem{twolevel-khrapai}
V.S. Khrapai, S. Ludwig, J.P. Kotthaus, H.P. Tranitz, and W. Wegscheider,
Phys. Rev. Lett. {\bf 97}, 176803 (2006).

\bibitem{twolevel-gustavsson}
S. Gustavsson, M. Studer, R. Leturcq, T. Ihn, K. Ensslin, D.C. Driscoll, A.C. Gossard,
Phys. Rev. Lett. {\bf 99}, 206804 (2007) .


\bibitem{qpc-field}
M. Field, C.G. Smith, M. Pepper, D.A. Ritchie, J.E.F. Frost, G.A.C. Jones, and D.G. Hasko,
Phys. Rev. Lett. {\bf 70}, 1311 (1993).

\bibitem{q-measurements1}
E. Buks, R. Schuster, M. Heiblum, D. Mahalu, and V. Umansky,
Nature (London) {\bf 391}, 871 (1998).


\bibitem{qpc-elzerman}
J.M. Elzerman, R. Hanson, J.S. Greidanus, L.H. Willems van Beveren,
S. De Franceschi, L.M.K. Vandersypen, S. Tarucha, and L.P. Kouwenhoven,
Phys. Rev. B {\bf 67}, 161308(R) (2003)

\bibitem{qpc-fujisawa}
T. Fujisawa, T. Hayashi, Y. Hirayama,
H.D. Cheong, and Y. H. Jeong,
Appl. Phys. Lett. {\bf 84}, 2343 (2004).

\bibitem{qpc-schleser}
R. Schleser, E. Ruh, T. Ihn, K. Ensslin
D.C. Driscoll, and A.C. Gossard,
Appl. Phys. Lett. {\bf 85},
2005 (2004).

\bibitem{qpc-vandersypen}
L.M.K. Vandersypen, J.M. Elzerman, R.N. Schouten, L.H. Willems van Beveren,
R. Hanson, and L.P. Kouwenhoven, Appl. Phys. Lett. {\bf 85}, 4394 (2004).

\bibitem{charge-dicarlo}
L. DiCarlo, H.J. Lynch, A.C. Johnson, L.I. Childress, K. Crockett, C.M. Marcus,
M.P. Hanson, and A.C. Gossard,
Phys. Rev. Lett. {\bf 92}, 226801 (2004).

\bibitem{charge-fujisawa}
T. Fujisawa, T. Hayashi, R. Tomita, and Y. Hirayama,
Science {\bf 312}, 1634 (2006).


\bibitem{sukhorukov1}
E.V. Sukhorukov, A.N. Jordan, S. Gustavsson, R. Leturcq, Th. Ihn, and K. Ensslin,
Nature Physics {\bf 3}, 243 (2007).


\bibitem{q-measurements5}
S.A. Gurvitz,
Phys. Rev. B {\bf 56}, 15215 (1997).

\bibitem{q-measurements7}
Y. Levinson,
Europhys. Lett. {\bf 39}, 299 (1997).

\bibitem{q-measurements8}
L. Stodolsky,
Phys. Rep. {\bf 320}, 51 (1999).


\bibitem{feedback-ensslin}
S. Gustavsson, R. Leturcq, M. Studer, I. Shorubalko, T. Ihn, K. Ensslin, D. C. Driscoll, A. C. Gossard
Surface Science Reports 64, 191 (2009)



\bibitem{Reznikov2}
M. Reznikov, M. Heiblum, H. Shtrikman, and D. Mahalu,
Phys. Rev. Lett. {\bf 75}, 3340 (1995).


\bibitem{Glattli}
A. Kumar, L. Saminadayar, D. C. Glattli, Y. Jin, and B. Etienne,
Phys. Rev. Lett. {\bf 76}, 2778 (1996).


\bibitem{position1}
C. B. Doiron, B. Trauzettel, and C. Bruder,
Phys. Rev. B {\bf 76}, 195312 (2007)


\bibitem{crosscorr1}
B. K\"{u}ng, O. Pf\"{a}ffli, S. Gustavsson, T. Ihn, K. Ensslin, M. Reinwald, W. Wegscheider
Phys. Rev. B 79, 035314 (2009)

\bibitem{crosscorr2}
U. Gasser, S. Gustavsson, B. K\"{u}ng, K. Ensslin, T. Ihn, D.C. Driscoll, A.C. Gossard
Phys. Rev. B 79, 035303 (2009)

\bibitem{jordan-buttiker}
A.N. Jordan, M. B\"{u}ttiker,
Phys. Rev. Lett. {\bf 95}, 220401 (2007).


\bibitem{position2}
C. B. Doiron, B. Trauzettel, and C. Bruder,
Phys. Rev. Lett. {\bf 100}, 027202 (2008).

\bibitem{position3}
S. Walter, and B. Trauzettel,
Phys. Rev. B {\bf 83}, 155411 (2011).


\bibitem{ingold-nazarov}
G.-L. Ingold and Y.V. Nazarov, in {\em Single
Charge Tunneling}, edited by H. Grabert and M.H. Devoret
(Plenum, New York, 1992), Chap. 2.


\bibitem{footnote-beyond-gaussian}
For a generalization beyond Gaussian noise, see Ref.\ \onlinecite{sonin} and Ref.\ \onlinecite{Sukhorukov2}.


\bibitem{sonin}
E.B. Sonin,
Phys. Rev. Lett. {\bf 98}, 030601 (2007)


\bibitem{footnote-mathematica}
we used Wolfram Mathematica package for analytical computing


\bibitem{FCS-ES Jordan}
A.N. Jordan, E.V. Sukhorukov.
Phys. Rev. Lett. {\bf 93}, 260604 (2004).

\bibitem{footnote-left-right-eigenvectors}
Note, that since in general case the transition matrix $\hat{M}$ is not symmetric, the left and right eigenvectors, $\langle m_j|$ and $|m_j\rangle$, are not Hermitian conjugated.


\bibitem{Sukhorukov2}
E. V. Sukhorukov, and J. Edwards
Phys. Rev. B {\bf 78}, 035332 (2008); arXiv:cond-mat/0804.0812.


\bibitem{footnote-transitions-interpretation}
Note that elements of $ \delta\!\hat{M}$ acquire a simple physical interpretation. For example: $\delta M_{14},\delta M_{23},\delta M_{32},\delta M_{41}$ represent the rates of correlated transitions $|11\rangle \leftrightarrow |22\rangle$ and $|12\rangle \leftrightarrow |21\rangle$, while $\delta M_{12},\delta M_{13},\delta M_{21},\delta M_{31}$ characterize the asymmetry of transitions $|11\rangle \leftrightarrow |21\rangle$ and $|12\rangle \leftrightarrow |22\rangle$, or, in other words, how the transition rates in one detector depend on the state of another.


\bibitem{footnote-slow-circuit}
For a general situation see the analysis in Ref.\ [\onlinecite{Sukhorukov2}].


\end{thebibliography}
\end{document}